# Two-Step Spike Encoding Scheme and Architecture for Highly Sparse Spiking-Neural-Network

Sangyeob Kim, *Graduate Student Member, IEEE*, Sangjin Kim, *Graduate Student Member, IEEE*,
Soyeon Um, *Graduate Student Member, IEEE*, Soyeon Kim, *Graduate Student Member, IEEE*,
and Hoi-Jun Yoo, *Fellow, IEEE*

*Abstract*— This paper proposes a two-step spike encoding scheme, which consists of the source encoding and the process encoding for a high energy-efficient spiking-neural-network (SNN) acceleration. The eigen-train generation and its superposition generate spike trains which show high accuracy with low spike ratio. Sparsity boosting (SB) and spike generation skipping (SGS) reduce the amount of operations for SNN. Time shrinking multi-level encoding (TS-MLE) compresses the number of spikes in a train along time axis, and spike-level clock skipping (SLCS) decreases the processing time. Eigen-train generation achieves 90.3% accuracy, the same accuracy of CNN, under the condition of 4.18% spike ratio for CIFAR-10 classification. SB reduces spike ratio by 0.49× with only 0.1% accuracy loss, and the SGS reduces the spike ratio by 20.9% with 0.5% accuracy loss. TS-MLE and SLCS increases the throughput of SNN by 2.8× while decreasing the hardware resource for spike generator by 75% compared with previous generators.

## I. INTRODUCTION

SPARSITY is very important for improving the performance of the convolutional-neural-network (CNN) processor. [1]-[4] However, the input sparsity of CNN varies severely, and it is difficult to increase the sparsity artificially. Spiking-Neural -Network (SNN) converts the data into sparse spike trains because there are many near zero values in input features due to Gaussian distribution. [5] Therefore, SNN can guarantee a high level of input sparsity unlike CNN, and it is possible to achieve high energy efficiency when accelerating a practical neural network. In addition, the accuracy of SNN is almost the same as CNN by CNN-to-SNN conversion method. [6]-[8]

Fig. 1 shows conventional spike encoding methods [9]-[11] for SNN. Rate encoding [9] generates a different number of spikes according to the magnitude of the input value. This method generates a binary pattern by comparing input data with the random number generated from the pseudorandom number generator (PRNG) every time step. However, it is not suitable for a high energy-efficient SNN architecture because of the high-power consumption of the PRNG unit. In addition, CNN-to-SNN conversion [6] has been recently proposed for SNN inference with the same accuracy as CNN, but the previous encodings are not suitable for this. For conversion,

The authors are with the School of Electrical Engineering, Korea Advanced Institute of Science and Technology, Daejeon 34141, South Korea.

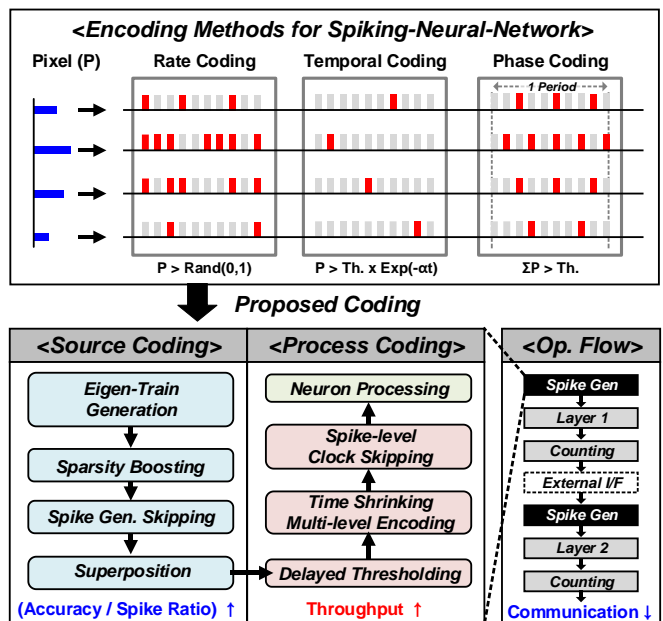

Fig. 1. Conventional and proposed spike coding method

pre-trained weights of CNN are shared with each neuron of the SNN, and each neuron sets a threshold using the received weights. The converted SNN can achieve a high accuracy of 69% for ImageNet [12] despite the 5 time-window condition [6]. The rate coding has randomness and generates different output spike train for the same value, which is different from the characteristics of CNN. Therefore, accuracy loss occurs when the converted weights are used. Time-to-first-spike (TTFS) coding [10] and phase coding [11] are also used for SNN with Spike-Time-Dependent-Plasticity (STDP) rule. TTFS generates only one spike for each input value, so it shows a low spike ratio. However, when TTFS is used with conversion to achieve high accuracy, it shows very low accuracy due to different characteristics of CNN. In CNN, multiple values are accumulated at the same time regardless of whether the values are large or small, but in TTFS, the large and small values are accumulated at different times. Phase coding continuously accumulates input value and generates a spike when it exceeds a specific value. In the process of comparing thresholds with large accumulation result, information from input value at low bit positions is lost, resulting in reduced accuracy. Therefore, in this article, we propose a new spike encoding method that can overcome the problems of the previous method for high accuracy and low spike ratio. In addition, we extend spike encoding method to two-step (source encoding and process coding), and suggest techniques to optimize each step. Source coding consists of 4 coding stages; Eigen-Train Generation (ETG), Sparsity Boosting (SB), Spike Generation Skipping (SGS), and Spike Train Superposition (STS). And, Process coding consists of 3 steps; Delayed Thresholding (DT), Time Shrinking Multi-Level Encoding (TS-MLE) and Spike-Level Clock Skipping (SLCS). After two step encoding, complete spike train with high sparsity is generated. Spike train is processed by SNN architecture, and we use counting and spike generation at every layer because it can reduce the communication by compressing the spike train.

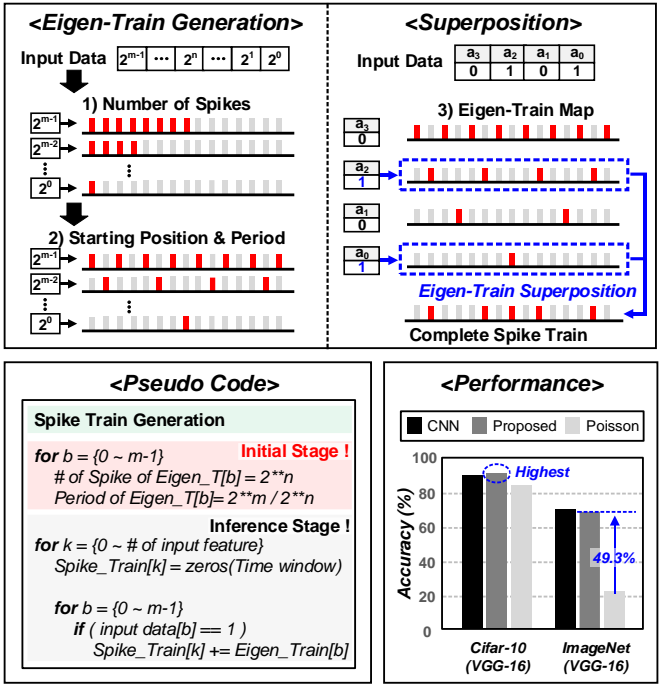

Fig. 2. Eigen-train generation and superposition for source encoding

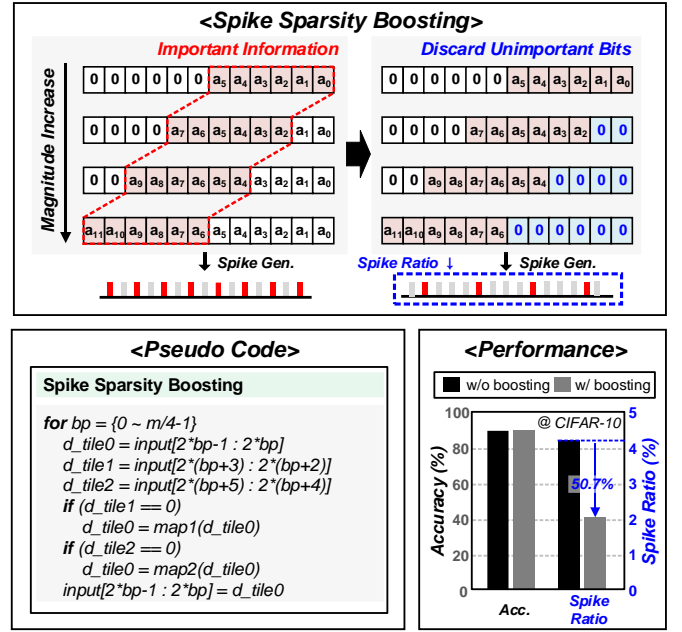

Fig. 3. Sparsity Boosting (SB) Method

## II. PROPOSED SPIKE ENCODING METHOD

### A. Source Encoding

Fig. 2 shows the ETG and STS. During the spike train generation process, a unique spike train is generated for each bit position to include information on all bits of data. The number of spikes and the period of each eigen-train are determined according to the position of the bit. Assuming that n is the bit position and m is the total number of bits in the data, the number of spikes corresponds to $2^n$ and the spike period corresponds to $2^m/2^n$. Then, some eigen-trains are superposed according to the input value. First, it is determined whether the value is 1 or 0 from the LSB position of the input value. If the corresponding value is 1, the eigen-train of the LSB is superposed to the spike train. This is repeated for the next bit position until the operation proceeds up to the MSB to generate the complete spike train. The proposed method maintains the information of all bit values during the spike train generation and excludes the randomness, so it shows higher accuracy than the previous encodings. For CIFAR-10 [13], the accuracy is 5.3% higher than the rate encoding. In the case of ImageNet, the proposed method achieves 49.3% higher accuracy.

Fig. 3 shows the SB technique to reduce the number of neuron operations. The larger the input value, the less important the information of the LSBs. Therefore, the number of spikes generated by LSB can be reduced with negligible accuracy loss unlike MSB. SB is performed before STS for m-bit input data. From the lowest 2 bits, it is checked whether the upper 2 bits are zero or not. If the upper 2 bits are zero, the value of the lowest 2 bits are maintained. Conversely, if the immediately preceding upper 2 bits are not zero, the lowest 2 bits are replaced with a value shifted by 1 bit to the right. In addition, it is checked whether the value is zero or not for the next high-order 2 bits. If the value is not zero, the lowest 2 bits are replaced with zero. After that, this process is repeated again from the second lowest 2 bits, and the repeating process is terminated when the position of m/2 bit is reached. In this way, unimportant ones of the original spikes are removed, increasing sparsity. Then, the data from SB is converted into a spike train through STS. As a result, the number of spikes is reduced by 50.7% under the condition of 0.1% accuracy loss for CIFAR-10 classification.

Fig. 4 shows the SGS that reduces the number of spikes and the amount of computation. In SNN, the membrane voltage of a neuron increases whenever it receives a spike from pre-neurons. When the membrane voltage exceeds a threshold, an output spike is generated. However, we propose a technique that predicts membrane voltage and omits the neuron operations since there are many neurons that may not generate any output. In general, neurons that do not generate output spike usually receive very few spikes from pre-neurons. Therefore, the number of output spikes from pre-neurons is counted to predict skipping. If the average value is below a threshold, the output spike of the corresponding neuron is predicted to be zero and recorded in the skipping map. After that, spike generation and neuron operation are omitted for neurons whose zero flag in the skipping map is "1". As a result, the spike ratio decreased by 20.9% with 0.5% accuracy loss for CIFAR-10 classification.

### B. Process Encoding

Fig. 5 shows the proposed SNN operation pattern and process coding. The operation pattern of conventional SNN [14], [15] is very inefficient in terms of weight reuse. The SNN operation is time serially processed, and the membrane voltage increases by the weight of each channel depending on whether a spike of each channel occurs at certain time step. At this time, weights should be fetched from weight memory serially, and the accumulated membrane voltage is compared with the threshold value of the neuron to determine whether an output spike fires. Thereafter, time step increases and a new membrane voltage is obtained using the newly received spikes. It requires weight data from weight memory again, and this operation is repeated until the time window size. During this operation, the previously loaded weights must be reloaded in order at every time step. Therefore, large memory power consumption is required because it cannot give chance to reuse weights in the direction of the time axis. The conventional method compares the membrane voltage with the threshold at every time step, but we propose a new method. In DT, the weight is just accumulated without thresholding ($\theta$ = threshold) for n time steps, and then, thresholding is performed with a new threshold $\theta_{DT}$ for the membrane voltage at the n-th time step. The example in Fig. 5 skips the thresholding for 4 time steps, and the new threshold $\theta_{DT}$ is $4\theta$. The $\theta_{DT}$ is a hyper parameter and its exact value can be determined at the training stage. DT can reuse the weight by n times after fetching, which reduces the power consumption required to access the weight memory. The number of reuses of the weights can also be greatly increased due to the proposed SNN operation pattern. As a result, it is possible to increase the weight reusability by 8 times for CIFAR-10 classification within a 0.2% accuracy drop.

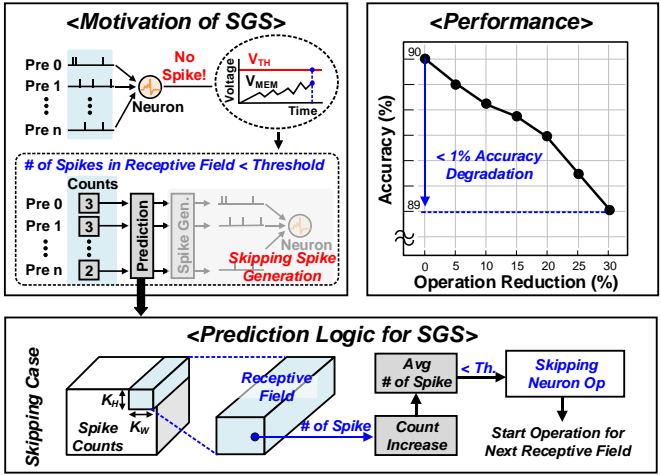

Fig. 4. Spike generation skipping (SGS) method

In addition, this paper proposes a TS-MLE and SLCS to increase the throughput. Most of the spike trains generated through spike encoding are very sparse. However, the conventional SNN consumes one cycle by one neuron even though it does not receive any spike. TS-MLE compresses the sparse spike train along the time axis to increase the hardware utilization. Bottom of Fig. 5 shows the example of TS-MLE and SLCS. When the spike train is sparse, 8 time-steps are compressed into 1 time-step as spikes with 4 levels. When the spike train is dense, it is compressed into a train of 2 time-steps of 4 level spikes. The input feature map of SNN has a Gaussian distribution, and most of the values are located near zero, resulting in a spike train with a small number of spikes. Therefore, spike trains compressed by TS-MLE also have small

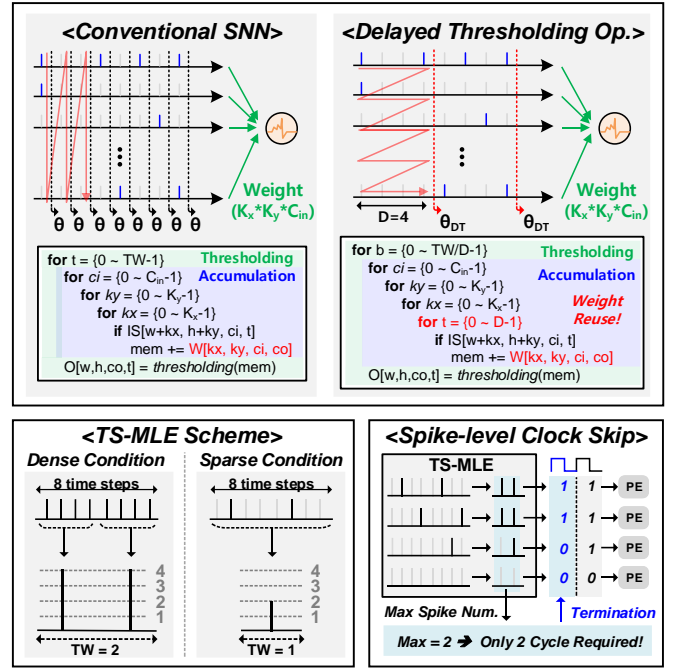

TW: Time Window, Cin: Input Channel, Ky: Kernel Width, Kx: Kernel Height, IS: Input Spike
Fig. 5. Process coding method

numbers of spikes with high probability. The SLCS can increase the throughput by skipping the operations of spike values smaller than maximum value of spike-level. Each processing element (PE) in SNN accelerator repeats the accumulation as much as the input multi-level spike value. Each PE receives spikes with a level of 0 to 3, and accordingly, the PE takes up to 3 cycles for operation. However, the spike ratio becomes very low by TS-MLE, and the multi-level spike that PEs receive may have less than the maximum value of 3 very often. In this case, if the real value of the spike delivered to PEs is less than 3, the operation is terminated before 3 cycles, and the next multi-level spikes can be delivered to the PE. For example, if the maximum spike value delivered to PEs is 1, all PEs terminate the multi-level spike operation after 1 cycle and receive the next spike. As a result, the throughput increases by 2.8x through TS-MLE and SLCS for CIFAR-10 classification.

III. IMPLEMENTATION RESULTS

Fig. 6 shows the performance comparison of the proposed method and others. In order to improve the performance of the SNN accelerator, a spike encoding that can achieve high accuracy while reducing the number of spikes is required. In this article, the spike ratio and accuracy are measured for different encodings with the CNN-to-SNN conversion for high accuracy. And, the performances of them are compared using specific figure-of-merit (FOM: accuracy/spike ratio). In the case of TTFS, it shows very low accuracy when the converted weights are used because the characteristic of the operation is different from the CNN. For rate and phase coding, spike ratios of 4.03% and 5.26% are achieved for CIFAR-10, respectively, and the accuracy for CIFAR-10 classification are 85.0% and 87.1%, respectively. Neither method is suitable for achieving high accuracy because the accuracy of the baseline CNN is 90.2%. When the ETG and STS are used, the accuracy of 90.3% is achieved under the condition of 4.18% spike ratio.

Fig. 6. Performance of Proposed Spike Encoding Methods

**<Performance of Spike Encoding Methods>**

| Methods | Spike Ratio | Accuracy | FOM * |
|---|---|---|---|
| @ CIFAR-10, ResNet-12, Time-window (TW) = 16 | | | |
| Rate | 4.03% | 85.0% | 21.1 |
| TTFS | 3.36% | 61.4% | 18.3 |
| Phase | 5.29% | 87.1% | 16.4 |
| Eigen Train Generation (ETG) | 4.18% | 90.3% | 21.6 |
| ETS w/ Sparsity Boosting (SB) | 2.06% | 90.2% | 43.7 |
| ETS w/ SB and Spike Gen. Skipping | 1.63% | 89.7% | 55.1 |
| @ CIFAR-100, VGG16, Time-window (TW) = 64 | | | |
| Rate | 5.41% | 62.2% | 11.5 |
| TTFS | 2.17% | 21.1% | 9.6 |
| Phase | 4.12% | 65.4% | 15.9 |
| Eigen Train Generation (ETG) | 3.76% | 66.9% | 17.8 |
| ETS w/ Sparsity Boosting (SB) | 2.39% | 66.6% | 27.9 |
| ETS w/ SB and Spike Gen. Skipping | 1.83% | 65.8% | 35.8 |
| @ ImageNet, ResNet-50, Time-window (TW) = 64 | | | |
| Rate | 5.49% | 22.8% | 4.2 |
| TTFS | - | 0.0% | 0.0 |
| Phase | 4.88% | 54.2% | 11.1 |
| Eigen Train Generation (ETG) | 4.92% | 72.1% | 14.7 |
| ETS w/ Sparsity Boosting (SB) | 4.34% | 71.3% | 16.4 |
| ETS w/ SB and Spike Gen. Skipping | 3.95% | 70.2% | 17.8 |

*  FOM = Accuracy / Spike Ratio

Compared to the conventional methods, a low spike ratio is achieved without loss of accuracy and the highest FOM value is achieved. In this article, two methods are proposed to increase the FOM. When SB is used, the FOM is increased by 2.02 times with only 0.1% accuracy degradation. When the SB and SGS are used together, a FOM of 55.1 are achieved with 0.6% loss in accuracy. As a result, the proposed encoding shows 2.61 times higher FOM compared to the previous methods for CIFAR-10 classification. Moreover, the proposed method achieves 2.25 times higher FOM for CIFAR-100 and 1.61 times higher FOM for ImageNet. In addition, the performance of different networks for ImageNet classification is reported in Fig. 7.

Fig. 8 shows the architecture of the generator for 8-bit input. It has a total of 8 eigen-train generator (ETG), and each ETG has 8 1-bit registers and 8 1-bit OR logics. The 8-bit register stores a unique eigen-train, and it outputs an eigen-train according to whether a specific bit of the input is zero or not. Outputs from ETGs are transferred to bit-wise OR logic, and the final spike train is generated. This encoder generates spikes within 8 time-steps per cycle, and the throughput is 8 times faster than the previous works [16] that generate only 1 spike per cycle. At the bottom of Fig. 8, there is a resource comparison with previous works [16]. Since the proposed generator shows faster speed, a value normalized by throughput is used to compare. The phase coding shows the fewest resources among the previous methods, and it uses 8 1b-Mux, 8 1b-Reg, and 76 NAND logics. However, the proposed generator requires only 15 1b-OR logic and 8 1b-Reg, which consumes lower resources despite showing higher accuracy and lower spike ratio. Therefore, the proposed method is the most hardware-friendly method in designing SNN accelerator.

## IV. HARDWARE ARCHITECTURE

Fig. 9 shows the SNN architecture which support the proposed spike generation. The input memory in the SNN core

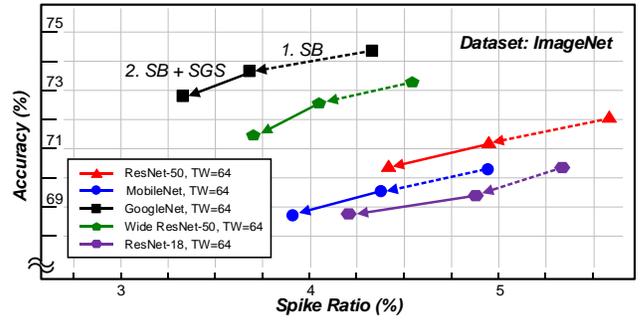

Fig. 7. Performance for Various Networks

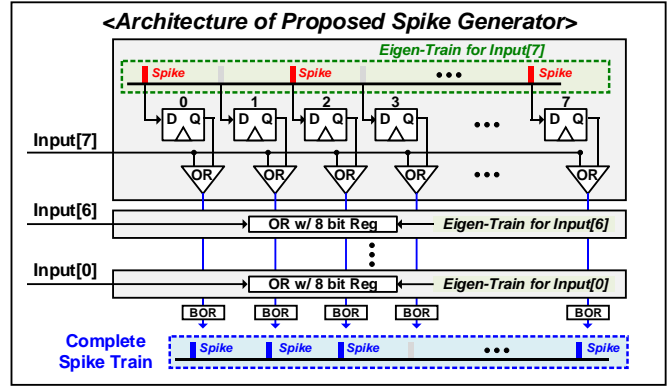

**<Hardware Resource of Spike Encoding Methods>**

| Gates | Rate | TTFS | Phase | Proposed |
|---|---|---|---|---|
| 1b-ADD | 0 | 12 | 0 | 0 |
| 1b-SUB | 0 | 0 | 0 | 0 |
| 8b-MUL | 0 | 1 | 0 | 0 |
| 1b-OR | 0 | 1 | 0 | 15 * |
| 1b-XOR | 3 | 0 | 0 | 0 |
| 1b-AND | 0 | 1 | 0 | 0 |
| 4b-Comp | 4 | 4 | 0 | 0 |
| 1b-MUX | 0 | 6 | 8 | 0 |
| 1b-Reg | 16 | 30 | 8 | 8 * |
| 40 x 9b-ROM | 0 | 2 | 0 | 0 |
| NAND | 316 | 2043 | 76 | 0 |

* Normalized by Throughput for Spike Generation

Fig. 8. Proposed Spike Generator Architecture and Performance

stores pixels or spike counts. When the operation starts, SGS unit controls the input memory. SGS unit has skipping map which includes the prediction results. If it is determined that the output spike is not zero, SGS unit transfers data request to the input memory. The SB unit performs input data converting to reduce the number of spikes. The converted data is transferred to the superposition unit, and the eigen-trains corresponding to the bit-position with a value of 1 are summed to generate a complete spike train. The generated spike train is transmitted to the TS-MLE unit, and it performs compression on the time axis to reduce the processing time. The compressed spike train is sent to the PE array, and each PE accumulates weights when it receives an input spike. After user-defined time-step, the partial sum is transferred to the thresholding unit, and it performs comparison and firing. In this case, multiple thresholding is required by delayed thresholding, and it is performed at once by dividing partial sum with threshold. Thereafter, the number of output spikes is counted in the accumulator of the thresholding unit, and it is stored in the output memory after time-window. As a result, the amount of operations and input parameters are reduced as shown in Fig. 10.

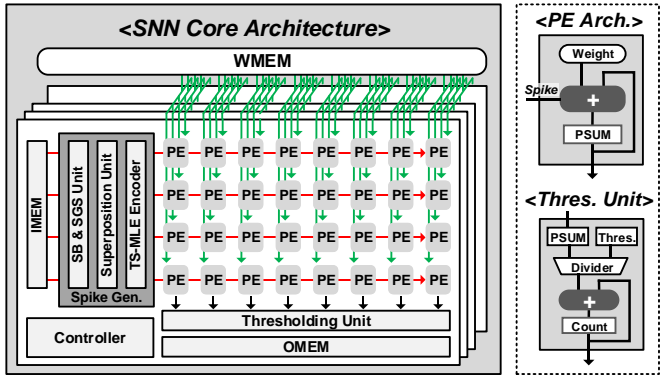

Fig. 9. SNN Architecture

**<Hardware Cost Comparison>**

| Methods | Computation (MOP) | Input Params (KB) | Accuracy (%) |
|---|---|---|---|
| @ CIFAR-10, ResNet-12, Time-window=16 | | | |
| CNN | 61 (x1.00) | 96 (x1.00) | 90.5 |
| SNN | 16 (x0.26) | 87 (x0.91) | 90.3 |
| SNN with Sparsity Boosting (SB) | 8 (x0.13) | 66 (x0.69) | 90.2 |
| SNN with SB and Spike Gen. Skipping | 6 (x0.09) | 54 (x0.56) | 89.7 |
| @ CIFAR-100, VGG16, Time-window=64 | | | |
| CNN | 761 (x1.00) | 221 (x1.00) | 67.2 |
| SNN | 463 (x0.61) | 203 (x0.92) | 66.9 |
| SNN with Sparsity Boosting (SB) | 292 (x0.38) | 141 (x0.64) | 66.6 |
| SNN with SB and Spike Gen. Skipping | 231 (x0.30) | 114 (x0.52) | 65.8 |
| @ ImageNet, ResNet-50, Time-window=64 | | | |
| CNN | 4972 (x1.00) | 5325 (x1.00) | 72.3 |
| SNN | 3021 (x0.61) | 4813 (x0.91) | 72.1 |
| SNN with Sparsity Boosting (SB) | 1915 (x0.39) | 4506 (x0.85) | 71.3 |
| SNN with SB and Spike Gen. Skipping | 1511 (x0.31) | 4096 (x0.77) | 70.2 |

Fig. 10. Hardware Cost Comparison

## V. CONCLUSIONS

In this paper, a hardware-friendly spike encoding is proposed. Previously, rate, TTFS, and phase coding have been proposed, but they cannot be used for a high-accuracy task due to information loss. In addition, methods such as rate encoding require a lot of resources because complex logic such as a PRNG. In this article, a two-step encoding scheme is proposed to solve these problems, and the source encoding and processing encoding are proposed. Source encoding includes ETG, STS, SB, and SGS. Through the STS, a spike ratio of 4.18% is achieved with an accuracy of 90.3% for CIFAR-10, resulting 0.77 times lower spike ratio. Also, SB achieves 2.07 times higher FOM compared to rate encoding, and the accuracy is reduced by only 0.1% compared to using ETS alone. When SB and SGS are used together, a 26% higher FOM is achieved with the 0.5% accuracy drop compared to only SB. Processing encoding includes DT, TS-MLE and SLCS. By using the compressed spike train, the processing time is reduced, and the throughput increased by 2.8 times for CIFAR-10. In addition, the proposed method requires the smallest amount of resources. When the sum of all logic is used as the FOM for hardware resource comparison, the proposed spike generation method requires 75% less resources than phase coding. As a result, the proposed encoding is the most suitable encoding method for SNN architecture because it shows the highest accuracy, the lowest spike ratio, and the lowest hardware resources.

REFERENCES

[1] J. Lee, S. Kang, J. Lee, D. Shin, D. Han and H. -J. Yoo, "The Hardware and Algorithm Co-Design for Energy-Efficient DNN Processor on Edge/Mobile Devices," in IEEE Transactions on Circuits and Systems I: Regular Papers, vol. 67, no. 10, pp. 3458-3470, Oct. 2020

[2] S. Kang, G. Park, S. Kim, S. Kim, D. Han and H. -J. Yoo, "An Overview of Sparsity Exploitation in CNNs for On-Device Intelligence With Software-Hardware Cross-Layer Optimizations," in IEEE Journal on Emerging and Selected Topics in Circuits and Systems, vol. 11, no. 4, pp. 634-648, Dec. 2021

[3] S. Kim, J. Lee, S. Kang, J. Lee and H. -J. Yoo, "A 146.52 TOPS/W Deep-Neural-Network Learning Processor with Stochastic Coarse-Fine Pruning and Adaptive Input/Output/Weight Skipping," 2020 IEEE Symposium on VLSI Circuits, 2020, pp. 1-2

[4] S. Kim, J. Lee, S. Kang, J. Lee, W. Jo and H. -J. Yoo, "PNPU: An Energy-Efficient Deep-Neural-Network Learning Processor With Stochastic Coarse–Fine Level Weight Pruning and Adaptive Input/Output/Weight Zero Skipping," in IEEE Solid-State Circuits Letters, vol. 4, pp. 22-25, 2021

[5] Sengupta, A., Ye, Y., Wang, R., Liu, C., and Roy, K. (2018). Going deeper in spiking neural networks: VGG and residual architectures. arXiv [Preprint]. arXiv:1802.02627.

[6] Nitin Rathi and Kaushik Roy. DIET-SNN: Direct Input Encoding With Leakage and Threshold Optimization in Deep Spiking Neural Networks. arXiv preprint arXiv:2008.03658, 2020.

[7] Wu, J., Chua, Y., Zhang, M., Li, G., Li, H., and Tan, K. C. (2019). A tandem learning rule for efficient and rapid inference on deep spiking neural networks. arXiv:1907.01167.

[8] J. Wu, C. Xu, D. Zhou, H. Li, and K. Chen Tan, "Progressive tandem learning for pattern recognition with deep spiking neural networks," 2020, arXiv:2007.01204.

[9] Srivastava, K. H., Holmes, C. M., Vellema, M., Pack, A. R., Elemans, C. P. H., Nemenman, I., et al. (2017). Motor control by precisely timed spike patterns. Proc. Natl. Acad. Sci. U.S.A. 114, 1171–1176. doi: 10.1073/pnas.1611734114

[10] Park, S., Kim, S. J., Na, B., and Yoon, S. (2020). "T2FSNN: deep spiking neural networks with time-to-first-spike coding," in Proceedings of the 2020 57th ACM/IEEE Design Automation Conference, San Francisco.

[11] Kim, J., Kim, H., Huh, S., Lee, J., and Choi, K. (2018). Deep neural networks with weighted spikes. Neurocomputing 311, 373–386. doi: 10.1016/j.neucom.2018.

[12] J. Deng, W. Dong, R. Socher, L. Li, Kai Li and Li Fei-Fei, "ImageNet: A large-scale hierarchical image database," 2009 IEEE Conference on Computer Vision and Pattern Recognition, Miami, FL, 2009, pp. 248-255

[13] Krizhevsky, A., & Hinton, G. (2009). Learning multiple layers of features from tiny images (Technical Report). University of Toronto.

[14] W. Guo, M. Fouda, A. Eltawil, and K. N. Salama. 2021. Neural Coding in Spiking Neural Networks: A Comparative Study for Robust Neuromorphic Systems. Frontiers in Neuroscience 15 (2021).

[15] G. K. Chen, R. Kumar, H. E. Sumbul, P. C. Knag and R. K. Krishnamurthy, "A 4096-Neuron 1M-Synapse 3.8-pJ/SOP Spiking Neural Network With On-Chip STDP Learning and Sparse Weights in 10-nm FinFET CMOS," in IEEE Journal of Solid-State Circuits, vol. 54, no. 4, pp. 992-1002, April 2019

[16] J. Park, J. Lee and D. Jeon, "A 65-nm Neuromorphic Image Classification Processor With Energy-Efficient Training Through Direct Spike-Only Feedback," in IEEE Journal of Solid-State Circuits, vol. 55, no. 1, pp. 108-119, Jan. 2020